\documentclass[reprint,nofootinbib,aps,prd,]{revtex4-2}

\usepackage{graphicx}
\usepackage{amsmath}
\usepackage{amssymb}
\usepackage{mathrsfs}
\usepackage{braket}
\usepackage{float}
\usepackage{bm}
\usepackage[colorlinks=true,citecolor=blue]{hyperref}

\newcommand{\be}{\begin{equation}}
\newcommand{\ee}{\end{equation}}
\newcommand{\bea}{\begin{eqnarray}}
\newcommand{\eea}{\end{eqnarray}}

\begin{document}

\title{A note on complete gauge-fixing and the constraint algebra}

\date{}
\author{Ganga Singh Manchanda}
\email{ganga@manchanda.co.uk}
\affiliation{}

\begin{abstract}
The admissibility of a gauge-fixing is governed by the invertibility of $\Delta=\{\sigma^a,\gamma_b\}$ where $\sigma^a$ are gauge-fixing conditions and $\gamma_b$ are independent first-class constraints. We prove, via the Schur complement, that the determinant of the combined constraint matrix $\mathcal{M}=\{\Phi_A, \Phi_B\}$ built from all constraints and gauge-fixing conditions factorises as $\det\mathcal{M}\approx\pm(\det\Delta)^2\det C$, where $C$ is the second-class constraint matrix, providing an alternative criterion for admissibility. Since $\det C\neq0$ by definition, the second-class sector decouples entirely from the gauge-fixing sector. In the algebraic case, this factorisation identifies the Hamiltonian admissibility criterion of Henneaux and Teitelboim~\cite{Henneaux} with the Lagrangian completeness criterion of Motohashi, Suyama, and Takahashi~\cite{Motohashi}. We identify a metric ansatz as gauge-fixing at the action level and analyse completeness in the context of spherically symmetric spacetime. The factorisation ensures that completeness is robust to the second-class sector that arises in modified
theories of gravity.
\end{abstract}

\maketitle

\section{Introduction}

Every gauge theory of physical interest is represented by a singular Lagrangian. This poses a problem because the Legendre transform fails to account for non-invertible conjugate momenta and thus neglects relations of the form, $\gamma_a(q,p)=0$, which cannot be inferred from a Legendre-transformed Hamiltonian but nevertheless express real physics. To pass between the Lagrangian and Hamiltonian forms of these theories requires the Dirac--Bergmann algorithm which identifies these constraints, adds them to the Hamiltonian by hand, classifies them, and ensures self-consistency~\cite{Dirac,Henneaux}.

For each independent first-class constraint, the corresponding Lagrange multiplier carries a gauge degree of freedom of the theory. The process of removing this freedom, gauge-fixing, is well understood at the Hamiltonian level. For $F$ independent first-class constraints we impose $F$ conditions, $\sigma^a\approx0$, which satisfy $\det\Delta\equiv\det\{\sigma^a,\gamma_b\}\neq0$. In systems with $S$ additional second-class constraints, $\chi_\alpha\approx 0$, practitioners often deal with the combined constraint set, $\Phi_A=(\gamma_a,\chi_\alpha,\sigma^a)$, which should be entirely second-class once gauge freedom is eliminated (i.e. $\det\mathcal{M}\equiv\det\{\Phi_A,\Phi_B\}\neq0$).

Naively, the admissibility criteria, $\det\Delta\neq0$ and $\det\mathcal{M}\neq0$, seem to be equivalent. This much is clear when no second-class constraints are present ($S = 0$), and the combined constraint matrix is the $2F\times2F$ block built from $\gamma_a$ and $\sigma^a$, with determinant $(\det\Delta)^2$. When second-class constraints are present, the combined constraint matrix is $(2F+S)\times(2F+S)$, and one might worry that cross-couplings between gauge conditions and second-class constraints, $R=\{\chi_\alpha,\sigma^b\}$, could spoil invertibility even when $\det\Delta\neq0$, or conversely rescue it when $\det\Delta=0$. We show that neither occurs. The determinant factorises exactly as:
\be\label{detM}
    \det\mathcal{M}\approx\pm(\det\Delta)^2\det C,
\ee
where $C=\{\chi_\alpha,\chi_\beta\}$. Since $\det C\neq0$ by definition, $\mathcal{M}$ is invertible if and only if $\Delta$ is invertible. The second-class sector decouples entirely from the gauge-fixing sector. Proof of~\eqref{detM} is simple via the Schur complement, and the result ensures that either $\det\mathcal{M}$ or $\det\Delta$ can verify the admissibility of a proposed gauge-fixing, with usage depending on context. In theories with complicated second-class sectors, this can provide a practical simplification.

The decoupling is in essence a consequence of the Dirac bracket construction and can be understood through that lens. Making the decoupling explicit at the level of a determinant identity has two advantages: it provides a self-contained criterion that does not require constructing the Dirac bracket, and it connects directly to the Lagrangian completeness criterion of Motohashi, Suyama, and Takahashi, which characterises when gauge functions are determined uniquely by gauge conditions~\cite{Motohashi}.

The plan of this paper is as follows. In \hyperref[sec:factorisation]{Section~\ref{sec:factorisation}} we establish some formal details of the constraint algebra and prove the relation between $\det\Delta$ and $\det\mathcal{M}$. \hyperref[sec:completeness]{Section~\ref{sec:completeness}} demonstrates the connection to the Lagrangian completeness criterion, and discusses consequences. Finally, in \hyperref[sec:application]{Section~\ref{sec:application}} we look at applications to gravity, diagnosing the completeness of a common spherically symmetric metric ansatz and noting the implications of our factorisation in modified theories of gravity where second-class constraints are generically present.

\section{Proof of factorisation}\label{sec:factorisation}

\subsection{Preliminaries}

Consider a constrained Hamiltonian system with $F$ first-class constraints, $\gamma_a\approx0$, and $S$ second-class constraints, $\chi_\alpha\approx0$. Before moving forward, we first set out some definitions and conditions which we will rely on.

We adopt the stronger definition of first-class as those constraints whose Poisson bracket with \emph{every} constraint in the theory vanishes weakly~\cite{Henneaux}. This can always be achieved by redefining $\gamma_a\to\gamma_a+k_a{}^\alpha \chi_\alpha$ with suitably chosen $k_a{}^\alpha$, such that $\Delta^a{}_b\to\Delta^a{}_b-k_b{}^\alpha R_\alpha{}^a$. While $\det\mathcal{M}$ and $\det C$ are intrinsic to the constraint surface, $\det\Delta$ is not; its value changes with the redefinition. The vanishing of $\det\Delta$ is, however, invariant, so that the conclusion $\det\Delta\neq0\Leftrightarrow\det\mathcal{M}\neq0$ holds regardless of the basis choice (under the strong definition of first-class).

All equalities are weak, holding only on the constraint surface. In particular, the Schur complement manipulations require $\det\Delta\neq0$ to hold on the constraint surface; if $\det\Delta$ vanishes at isolated points of this surface, the factorisation and the admissibility conclusion hold only away from those points. At isolated points where $\det\Delta=0$, the gauge-fixing conditions fail, corresponding to a Gribov-type obstruction (gauge slices are tangent to orbits)~\cite{Gribov}.

Finally, we note that $\det C\neq0$ on the surface $\{\gamma_a=0,\chi_\alpha=0\}$ implies $\det C\neq0$ on the surface $\{\gamma_a=0,\chi_\alpha=0,\sigma^a=0\}$, since $\det C$ is a smooth non-vanishing function on the larger surface. The combined constraint set is then entirely second-class on the full constraint surface when $\det\mathcal{M}\neq0$, which the factorisation relates to $\det\Delta\neq0$. With these conventions in place, we turn to the proof.

\subsection{$\det\Delta\neq0\Longrightarrow\det\mathcal{M}\neq0$}

We permute the combined constraints as $\tilde\Phi_A=(\gamma_a,\sigma^a,\chi_\alpha)$ such that the permuted $\det\tilde{\mathcal{M}}$ differs from $\det\mathcal{M}$ only up to a sign. On the constraint surface the combined constraint matrix evaluates to:
\be\label{Mtilde}
    \tilde{\mathcal{M}}\approx
    \begin{pmatrix}
    0 & -\Delta^T & 0 \\
    \Delta & Q & -R^T \\
    0 & R & C
    \end{pmatrix},
\ee
where $\Delta=\{\sigma^a,\gamma_b\}$, $Q=\{\sigma^a, \sigma^b\}$, $R=\{\chi_\alpha,\sigma^b\}$, and $C=\{\chi_\alpha,\chi_\beta\}$. The zeroes in~\eqref{Mtilde} follow from the stronger definition of first-class constraints. We label the upper-left $2F\times2F$ block:
\be
    \mathcal{A} \equiv
    \begin{pmatrix}
    0 & -\Delta^T \\
    \Delta & Q
    \end{pmatrix}.
\ee
When $\det\Delta\neq0$, this block has $\det\mathcal{A}=(\det\Delta)^2$, and inverse:
\be
    \mathcal{A}^{-1}=
    \begin{pmatrix}
    \Delta^{-1}Q\Delta^{-T} & \Delta^{-1}\\
    -\Delta^{-T} & 0
    \end{pmatrix}.
\ee
The Schur complement of $\mathcal{A}$ in $\tilde{\mathcal{M}}$~\cite{Crabtree}:
\be
    \tilde{\mathcal{M}}/\mathcal{A}=C-
    \begin{pmatrix} 0 & R \end{pmatrix}
    \mathcal{A}^{-1}
    \begin{pmatrix} 0 \\ -R^T \end{pmatrix}=C,
\ee
demonstrates that the cross-couplings, $R$, drop out entirely such that:
\be
    \det\tilde{\mathcal{M}}\approx\det\mathcal{A}\det C=(\det\Delta)^2\det C.
\ee
Since $\det C\neq0$ by definition, $\det\mathcal{\tilde{M}}\neq0$ and $\det\mathcal{M}\neq0$.

\subsection{$\det\Delta=0\Longrightarrow\det\mathcal{M}=0$}

If $\det\Delta=0$, there exists some non-zero $v^a$ with $\Delta^b{}_av^a=0$. We can construct $V^A = (v^a,0^a,0_\alpha)$ indexed over $(\gamma_a,\sigma^a,\chi_\alpha)$, and column-by-column, we see:
\begin{align*}
\gamma\text{ column:}\quad&v^a\{\gamma_a,\gamma_b\}=v^aU^c{}_{ab}\gamma_c\approx 0,\\
\sigma\text{ column:}\quad&-v^a(\Delta^T)^b{}_a=-\Delta^b{}_av^a=0,\\
\chi\text{ column:}\quad&v^a\{\gamma_a,\chi_\beta\}\approx0,
\end{align*}
such that $V^A\tilde{\mathcal{M}}_{AB}\approx0$. $\tilde{\mathcal{M}}$ is therefore singular on the constraint surface, implying $\det\mathcal{M}\approx0$.

\section{Connection to Lagrangian completeness}\label{sec:completeness}

In the Lagrangian framework, gauge-fixing at the action level (``pre-fixing'') is safe if it is complete in the sense of~\cite{Motohashi}: the conditions must satisfy $\delta_\varepsilon \sigma^a = 0$, in order to uniquely determine the gauge parameters $\varepsilon^a$. In general, $\delta_\varepsilon \sigma^a = M^{(0)a}{}_b\varepsilon^b + M^{(1)a}{}_b\dot\varepsilon^b + \cdots$, and completeness requires $M^{(0)a}{}_b$ to have trivial kernel.

The connection to the Hamiltonian admissibility criterion is direct in the algebraic case, where the gauge transformation of $\sigma^a$ involves the gauge parameters but not their time derivatives. Here the canonical gauge transformation reduces to $\delta_\varepsilon\sigma^a=\varepsilon^b\{\sigma^a,\gamma_b\}$, so that the zeroth-order coefficient $M^{(0)a}{}_b=\{\sigma^a,\gamma_b\}=\Delta^a{}_b$, and completeness is equivalent to $\det\Delta\neq0$. The factorisation~\eqref{detM} then implies that the criterion is unaffected by any second-class constraints in the theory, since the second-class sector contributes only the factor $\det C$, which is non-vanishing by definition. When derivative terms are present, $\det\Delta\neq0$ is neither sufficient nor necessary for completeness (see, e.g.~\cite{Motohashi}).

The factorisation, being an algebraic identity about Poisson bracket matrices, remains valid regardless; however, one should note a subtlety: if completeness holds with $\det\Delta=0$, then the factorisation implies $\det\mathcal{M}=0$. In this case, the combined constraint set, $\Phi_A$, is not entirely second-class at this stage, even though the gauge is fully fixed. This is not a contradiction, though, as the gauge-fixing conditions $\sigma^a$ are not yet ``final'' in the Hamiltonian sense. Requiring $\dot\sigma^a\approx0$ generates additional secondary conditions (since time preservation of the conditions cannot determine all multipliers when $\det\Delta=0$), and these new conditions complete the second-class set upon further iteration of the Dirac--Bergmann algorithm.

\section{Application to gravity}
\label{sec:application}

Working with a symmetric metric in the Hamiltonian formulation raises an immediate concern.  The most general spherically symmetric line element,
\be\label{eq:spherical}
    ds^2=-A\,dt^2+\frac{dr^2}{B}+2C\,dtdr+Er^2\gamma_{ij}\,dx^idx^j,
\ee
contains four free functions $(A,B,C,E)$. In practice, one almost always sets $C=0$ and $E=1$ before varying the action, treating this as an innocuous simplification. But a metric ansatz does two things at once. It restricts the spatial metric $h_{ij}$ to a form compatible with the assumed symmetry, a reduction justified by Palais' principle~\cite{Palais}. It also restricts the lapse and shift, which amounts to a gauge-fixing at the action level, subject to the completeness criterion. These independent operations are often conflated, leading to concerns about incomplete gauge-fixing. Palais' principle guarantees that the symmetry-reduction is safe, but says nothing about whether the simultaneous gauge choice preserves the correct number of independent equations.

A completeness analysis of the gauge conditions $C=0$, $E=1$ for the metric~\eqref{eq:spherical} shows that when the fields depend on both $t$ and $r$, enforcing the ansatz leaves the gauge parameter $\varepsilon^t(t)$ undetermined~\cite{Motohashi}. In this scenario, the completeness operator has non-trivial kernel and the gauge-fixing is incomplete, leading to an unphysical time-dependent mass in the Schwarzschild solution. If all fields were instead static (functions of $r$ only), the completeness matrix would be invertible and the gauge-fixing complete. 

The factorisation adds a further conclusion. Since vacuum ADM gravity has no second-class constraints ($S=0$), the combined constraint matrix gives $\det\mathcal{M}=(\det\Delta)^2$, and completeness is straightforward. In modified theories of gravity (e.g. scalar-tensor~\cite{Horndeski}, massive gravity~\cite{deRham}), however, second-class constraints are generically present. One might worry, in the same way as before, about cross-couplings, $R$, rendering the earlier conclusions unreliable in the modified setting. This does not occur, as the second-class sector contributes only the factor $\det C$, which is non-vanishing by definition. The completeness of the gauge-fixing is determined by $\det\Delta$ alone, regardless of the second-class content of the theory.

\section{Discussion}

The factorisation reads naturally at the level of Dirac brackets. Under the strong definition of first-class, the difference between Poisson and Dirac brackets (in the first-class sector) vanishes. The admissibility criterion is thus insensitive to the choice of bracket, and the factorisation is a determinant-level consequence of that insensitivity. The strong first-class definition allows for the factorisation algebraically but also means gauge orbits preserve the second-class surface, causing the two sectors to decouple geometrically.

Two limitations should be noted. First, the result assumes regularity of all constraints, such that theories with ineffective constraints lie outside its scope~\cite{Gotay}. Second, in the non-algebraic case where $\delta_{\varepsilon}\sigma^{a}$ involves time derivatives of the gauge parameters, Lagrangian completeness is controlled by a differential operator $\hat{P}^{a}{}_{b}$ whose zeroth-order part is $\Delta$~\cite{Motohashi}. The factorisation remains valid as an algebraic identity but does not by itself capture the full completeness condition, and whether a structurally analogous decoupling holds in this context remains open.

\section*{Acknowledgements}

The author acknowledges the assistance of Opus 4.6 in the preparation of this note and assumes full responsibility for its scientific content and integrity.


\begin{thebibliography}{99}

\bibitem{Henneaux}
M.~Henneaux and C.~Teitelboim,
\emph{Quantization of Gauge Systems} (Princeton University Press, Princeton, 1992).

\bibitem{Motohashi}
H.~Motohashi, T.~Suyama, and K.~Takahashi,
``Fundamental theorem on gauge fixing at the action level'',
\href{https://doi.org/10.1103/PhysRevD.94.124021}{{\it Phys. Rev. D} \textbf{94}, 124021 (2016)},
\href{https://doi.org/10.48550/arXiv.1608.00071}{[arXiv:1608.00071 [gr-qc]]}.

\bibitem{Dirac}
P.~A.~M.~Dirac,
\emph{Lectures on Quantum Mechanics} (Yeshiva University, New York, 1964).

\bibitem{Gribov}
V.~N.~Gribov,
``Quantization of non-Abelian gauge theories'',
\href{https://doi.org/10.1016/0550-3213(78)90175-X}{{\it Nucl. Phys. B} \textbf{139}, 1--19 (1978)}.

\bibitem{Crabtree}
D.~E.~Crabtree and E.~V.~Haynsworth,
``An identity for the Schur complement of a matrix'',
\href{https://doi.org/10.1090/S0002-9939-1969-0255573-1}{{\it Proc. Amer. Math. Soc.} \textbf{22}, 364--366 (1969)}.

\bibitem{Palais}
R.~S.~Palais,
``The principle of symmetric criticality'',
\href{https://doi.org/10.1007/BF01941322}{{\it Commun. Math. Phys.} \textbf{69}, 19 (1979)}.

\bibitem{Horndeski}
G.~W.~Horndeski,
``Second-order scalar-tensor field equations in a four-dimensional space'',
\href{https://doi.org/10.1007/BF01807638}{{\it Int. J. Theor. Phys.} \textbf{10}, 363--384 (1974)}.

\bibitem{deRham}
C.~de Rham, G.~Gabadadze, and A.~J.~Tolley,
``Resummation of Massive Gravity'',
\href{https://doi.org/10.1103/PhysRevLett.106.231101}{{\it Phys. Rev. Lett.} \textbf{106}, 231101 (2011)},
\href{https://doi.org/10.48550/arXiv.1011.1232}{[arXiv:1011.1232 [hep-th]]}.

\bibitem{Gotay}
M.~J.~Gotay,
``On the validity of Dirac's conjecture regarding first-class secondary constraints'',
\href{https://doi.org/10.1088/0305-4470/16/5/003}{{\it J. Phys. A: Math. Gen.} \textbf{16}, L141 (1983)}.

\end{thebibliography}
\end{document}